\documentclass[11pt]{article}
\usepackage[dvips]{graphicx}
\usepackage{mathrsfs}
\usepackage{amsthm} 
\usepackage{amsmath}
\usepackage{amssymb}
\usepackage{amsfonts}
\usepackage{float}
\usepackage{yhmath}
\usepackage{wrapfig}
\usepackage[dvips]{color}
\usepackage[dvips]{colortbl}
\usepackage{cite}
\usepackage{array}
\usepackage{tabularx}
\usepackage[sectionbib]{chapterbib}
\usepackage{threeparttable}
\usepackage{chemarrow}
\usepackage{framed}
\usepackage{ascmac}
\definecolor{shadecolor}{gray}{0.90}
\usepackage[small, bf, labelsep=quad]{caption}
\usepackage{cases}

\newcommand{\II}{I\hspace{-0.5mm}I}

\definecolor{mycolor1}{rgb}{1,1,0.7}
\definecolor{mycolor2}{rgb}{0.9,1,1}
\definecolor{mycolor3}{cmyk}{0,0,0,0.113}
\definecolor{mycolor4}{cmyk}{0.086,0,0,0}

\begin{document}

\begin{center}
\Huge{Branched Polymers with Excluded Volume Effects}
\end{center}
\vspace*{-3mm}
\begin{center}
\LARGE{Configurations of Comb Polymers in Two- and Three-dimensions}
\end{center}

\vspace*{10mm}
\begin{center}
\large{Kazumi Suematsu\footnote{\, The author takes full responsibility for this article.}, Haruo Ogura$^{2}$, Seiichi Inayama$^{3}$, and Toshihiko Okamoto$^{4}$} \vspace*{2mm}\\
\normalsize{\setlength{\baselineskip}{12pt} 
$^{1}$ Institute of Mathematical Science\\
Ohkadai 2-31-9, Yokkaichi, Mie 512-1216, JAPAN\\
E-Mail: suematsu@m3.cty-net.ne.jp, ksuematsu@icloud.com  Tel/Fax: +81 (0) 593 26 8052}\\[3mm]
$^{2}$ Kitasato University,\,\, $^{3}$ Keio University,\,\, $^{4}$ Tokyo University\\[15mm]
\end{center}

\hrule
\vspace{3mm}
\noindent
\textbf{\large Abstract}: We investigate the excluded volume effects in good solvents for the isolated comb polymers having $\nu_{0}=1/4$. In particular, we investigate the change of the size exponent, $\nu$, defined by $\langle s_{N}^{2}\rangle\propto N^{2\nu}$, for the various fully-expanded configurations. The results show that, given the fully-stretched backbone and side chains, the exponent takes the value, $\nu=1/2$, irrespective of the configurational isomerization of side chains; only the pre-exponential factor changes. 

\vspace{-2mm}
\begin{flushleft}
\normalsize{}\\[3mm]
\end{flushleft}
\hrule
\vspace{10mm}
\setlength{\baselineskip}{14pt}

\section{Introduction}
Until recently, only a few ideal size-exponents, $\nu_{0}$, had been discovered for polymeric materials: (i) $\nu_{0}=1/2$ for linear molecules, (ii) $1/4$ for randomly branched polymers, and (iii) $0$ for dendrimers. In the preceding works\cite{Kazumi2}, we put forth that an infinite number of architectures with different exponents can be constructed based on the nesting procedure. In the process of evaluating the configurational characteristic of each architecture, it was suggested that there might exist an empirical rule that satisfies
\begin{equation}
\langle s_{g}^{2}\rangle=
\begin{cases}
\hspace{1mm}A\hspace{0.3mm}g\hspace{0.3mm}l^{2} & \mbox{(for freely jointed molecules)}\\[3mm]
\hspace{1mm}A'g^{2}l^{2}& \mbox{(for fully expanded molecules)}
\end{cases}\label{gz2poly-1}
\end{equation}
where $\langle s_{g}^{2}\rangle$ is the mean square of the radius of gyration, $A$ and $A'$ are coefficients that change depending on architectures, and $g$ the generation number from the root to the youngest (outermost) generation on the main backbone. Given Eq. (\ref{gz2poly-1}), it makes us possible, with the help of the relationship between $g$ and $N$, to calculate directly the size exponent, $\nu$, defined by $\langle s_{N}^{2}\rangle\propto N^{2\nu}$\cite{Flory, Issacson, Seitz, Daoud, Kazumi2, Rosa}. As is listed in Table \ref{MeanRGFreelyJointed}, the first relation, $\langle s_{g}^{2}\rangle\propto g$, in Eq. (\ref{gz2poly-1}) has been confirmed for all the known freely-jointed architectures. So, it appears that the first equality may be considered to be a theorem rather than an empirical equation. The main aim of this paper is, thus, concentrated on the examination of the second relation, $\langle s_{g}^{2}\rangle\propto g^{2}$.

In the present examination, we make full use of the information on the regular comb polymer having side chains of the length, $n$. This is because this polymer has a versatile configurational backbone suitable to develop our arguments.

 \begin{table}[h]
 \centering
  \begin{threeparttable}
    \caption{The asymptotic formulae for the mean squares of the radii of gyration for architectures without excluded volume.}\label{MeanRGFreelyJointed}
\vspace*{-2mm}
  \begin{tabular}{l r l l}
\hline\\[-2mm]
\hspace{5mm} polymer  \hspace{10mm} &formulae for $\left\langle s_{N}^{2}\right\rangle_{0}$ &\hspace{4mm} relationship between $N$ and $g$ &\hspace{3mm}  $\nu_{0}$\,\,\,\,\\[2mm]
\hline\\[-1.5mm]
\hspace{5mm} linear  & $\big(\frac{1}{6}\big)\hspace{0.3mm}g\hspace{0.4mm}l^{2}$\hspace{7 mm} &\hspace{4mm}  $N=g$ &\hspace{3mm}  $\frac{1}{2}$\\[1.5mm]
\hspace{5mm} star (equal arm length)\tnote{\,a}  &$\big(\frac{3f-2}{6f}\big)\hspace{0.3mm}g\hspace{0.4mm}l^{2}$\hspace{7 mm} &\hspace{4mm}  $N=1+f(g-1)$ &\hspace{3mm}  $\frac{1}{2}$\\[1.5mm]
\hspace{5mm} regular comb  &$\big(\frac{1}{6}\big)\hspace{0.3mm}g\hspace{0.4mm}l^{2}$\hspace{7 mm} &\hspace{4mm}  $N=g+(g-1)n$ &\hspace{3mm}  $\frac{1}{2}$\\[1.5mm]
\hspace{5mm} extended comb\tnote{\,b}  &$\big(\frac{2}{3}\big)\hspace{0.3mm}g\hspace{0.4mm}l^{2}$\hspace{7 mm} &\hspace{4mm}  $N=g^{2}$ &\hspace{3mm}  $\frac{1}{4}$\\[1.5mm]
\hspace{5mm} triangular\tnote{\,c}  &$\big(\frac{7}{15}\big)\hspace{0.3mm}g\hspace{0.4mm}l^{2}$\hspace{7 mm} &\hspace{4mm}  $N=g+\frac{1}{2}(g-1)(g-2)$ &\hspace{3mm}  $\frac{1}{4}$\\[1.5mm]
\hspace{5mm} randomly branched  &\rule[2pt]{10mm}{0.5pt}\hspace{10 mm} &\hspace{6mm}\rule[2pt]{1cm}{0.5pt} &\hspace{3mm}  $\frac{1}{4}$\\[1.5mm]
\hspace{5mm} $z$=3\tnote{\,d}  &$\big(\frac{7}{6}\big)\hspace{0.3mm}g\hspace{0.4mm}l^{2}$\hspace{7 mm} &\hspace{4mm}  $N=g(g^{2}-g+1)$ &\hspace{3mm}  $\frac{1}{6}$\\[1.5mm]
\hspace{5mm} dendrimers\tnote{\,e} &$g\hspace{0.4mm}l^{2}$\hspace{7 mm} &\hspace{4mm}  $N=\frac{f-3+(f-1)^{g-1}}{f-2}$ &\hspace{3mm}  0\\[1.5mm]
\hline\\[-6mm]
   \end{tabular}
   \vspace*{1mm}
   \begin{tablenotes}
     \item a. the star polymer with an equal arm length\cite{Kazumi2}; b. $n=g$; c. all the end monomers terminate at the $g$th generation; d. the nested architecture with the depth $z$=3\cite{Kazumi2}; e. \cite{Kazumi2, Polinska, Yang}.
   \end{tablenotes}
  \end{threeparttable}
  \vspace*{5mm}
\end{table}

\section{Configuration of the Comb Polymer in the Two-dimensional Space}\label{CCP2D}
\begin{figure}[H]
\vspace*{-1mm}
\begin{center}
\includegraphics[width=12cm]{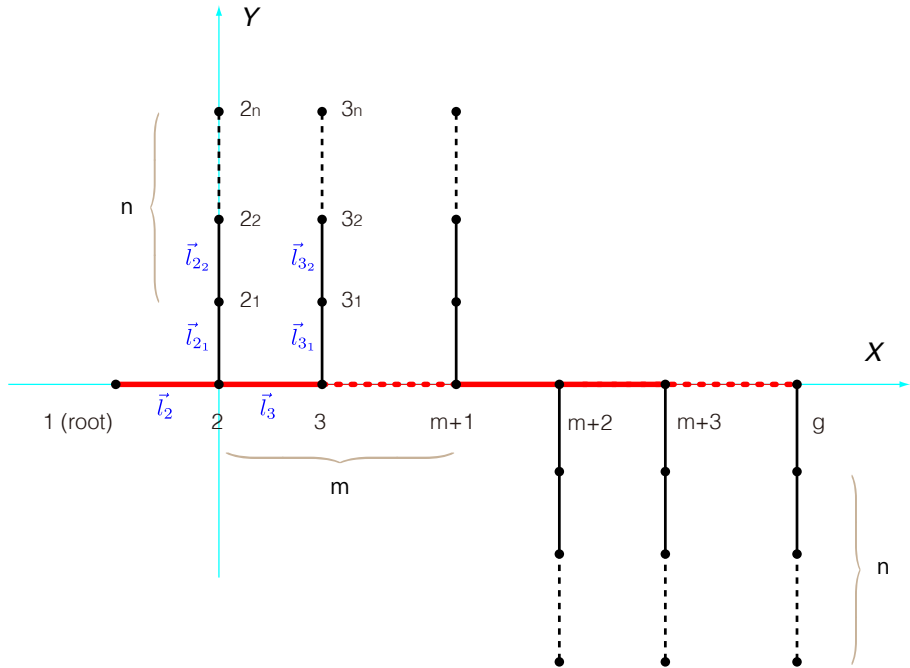} 
\caption{The comb polymer having the stretched backbone and the stretched side chains on the square lattice, with $m$ side chains extending to $Y(+)$ and $g-1-m$ to $Y(-)$.}\label{2D-z2}
\end{center}
\end{figure}

Let us consider the lattice analogue of the regular comb polymer on the square lattice. Let the cluster have a main backbone comprised of $g$ monomers and $g-1$ side chains with the length, $n$. Suppose this cluster is in the fully expanded state so that the main backbone (the red solid-line in Fig. \ref{2D-z2}) and the side chains have the stretched configuration.
What we are going to investigate are the spatial configurations of the cluster in which $m$ side chains extend to the $Y(+)$ direction and the remaining $g-1-m$ side chains to the opposite direction, $Y(-)$, as illustrated in Fig. \ref{2D-z2}. For such generalized configurations, we want to calculate the change of the mean radius of gyration as a function of $m$; this can be accomplished with the help of the Isihara formula\cite{Isihara, Kazumi2}:
\begin{equation}
\vec{r}_{Gp}=\vec{r}_{1p}-\frac{1}{N}\sum_{p=1}^{N}\vec{r}_{1p}\label{gz2poly-2}
\end{equation}
Note that the vector, $\vec{r}_{1p}$, can be expressed as the sum of bond vectors lying between the monomer 1 and $p$. So, the end-to-end vector, $\vec{r}_{Gp}$, from the center of mass to the $p$th monomer can be recast into the grand sum of all the bond vectors that constitute the polymer\cite{Isihara, Weiss, Redner, Kazumi2}. Following the definition introduced in the preceding paper\cite{Kazumi2}, let each side chain be a part of the corresponding monomer on the main backbone. We index the branching units on the backbone from 1 to $g$, and the units on the side chains from 1 to $n$; for instance, $3_{2}$ denotes the second monomers on the side chain emanating from the branching unit of the third generation (Fig. \ref{2D-z2}). The result is, for $p=h$,
\begin{description}
\item[]
\begin{multline}
\vec{r}_{Gh}=\frac{1}{N}\left\{\sum_{k=1}^{h-1}[N-(N-k-(k-1)n)]\,\vec{l}_{(k+1)}-\sum_{k=h}^{g-1}\left[N-k-(k-1)n\right]\,\vec{l}_{(k+1)}\right.\\
\left.-\sum_{k=1}^{m}\sum_{i=1}^{n}(n-i+1)\,\vec{l}_{(k+1)_{i}}+\sum_{k=m+1}^{g-1}\sum_{i=1}^{n}(n-i+1)\,\vec{l}_{(k+1)_{i}}\right\}\label{gz2poly-3}
\end{multline}
for $1\le h\le g$.
\begin{multline}
\hspace{-0.5cm}\vec{r}_{Gh_{j}}^{\hspace{3mm}+}=\frac{1}{N}\left\{\sum_{k=1}^{h-1}[N-(N-k-(k-1)n)]\,\vec{l}_{(k+1)}-\sum_{k=h}^{g-1}\left[N-k-(k-1)n\right]\,\vec{l}_{(k+1)}\right.\\
+\sum_{i=1}^{j}[N-n+i-1]\,\vec{l}_{h_{i}}-\sum_{k=1}^{m}\sum_{i=1}^{n}(n-i+1)\,\vec{l}_{(k+1)_{i}}\\
\left.+\sum_{k=m+1}^{g-1}\sum_{i=1}^{n}(n-i+1)\,\vec{l}_{(k+1)_{i}}+\sum_{i=1}^{j}[n-i+1]\,\vec{l}_{h_{i}}\right\}\label{gz2poly-4}
\end{multline}
for $2\le h\le m+1$.
\begin{multline}
\hspace{-0.5cm}\vec{r}_{Gh_{j}}^{\hspace{3mm}-}=\frac{1}{N}\left\{\sum_{k=1}^{h-1}[N-(N-k-(k-1)n)]\,\vec{l}_{(k+1)}-\sum_{k=h}^{g-1}\left[N-k-(k-1)n\right]\,\vec{l}_{(k+1)}\right.\\
-\sum_{i=1}^{j}[N-n+i-1]\,\vec{l}_{h_{i}}-\sum_{k=1}^{m}\sum_{i=1}^{n}(n-i+1)\,\vec{l}_{(k+1)_{i}}\\
\left.+\sum_{k=m+1}^{g-1}\sum_{i=1}^{n}(n-i+1)\,\vec{l}_{(k+1)_{i}}-\sum_{i=1}^{j}[n-i+1]\,\vec{l}_{h_{i}}\right\}\label{gz2poly-5}
\end{multline}
for $m+2\le h\le g$.
\end{description}
The key point in Eqs. (\ref{gz2poly-3})-(\ref{gz2poly-5}) is that the direction of the bond vectors, $\vec{l}_{(k+1)_{i}}$, on the side chains is \textit{plus} [$Y(+)$] for $k=1$ to $m$ and \textit{minus} [$Y(-)$] for $k=m+1$ to $g-1$.

\begin{figure}[h]
\begin{center}
\includegraphics[width=9cm]{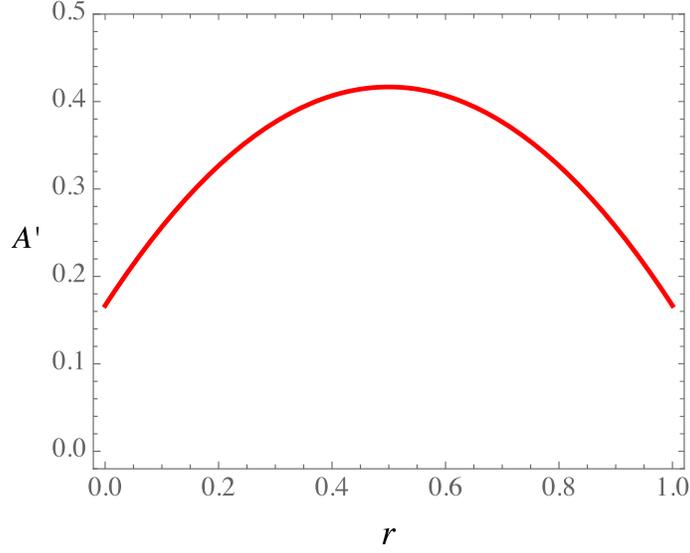} 
\caption{The prefactor, $A'=\frac{1}{6}(1+6r-6r^{2})$, as against the ratio, $r=\frac{m}{g-1}$, for the comb polymer having the stretched backbone and the stretched side chains on the square lattice, with $m$ side chains extending to $Y(+)$ and $g-1-m$ to $Y(-)$ (see Fig. \ref{2D-z2}).}\label{FullyExpanded-B}
\end{center}
\vspace*{-4mm}
\end{figure}

Note that the main backbone and the side chains extend perpendicularly to each other, so that $\langle\vec{l}_{h}\cdot\vec{l}_{h_{i}}\rangle=0$ ($h=2, 3, \cdots, g$). Let all bonds have the same length, $|\vec{l}|=l$. The mean squares of the end-to-end distance are
\begin{description}
\item[]
\begin{multline}
\left\langle r_{Gh}^{2}\right\rangle=\frac{l^{2}}{N^{2}}\left\{\left(\sum_{k=1}^{h-1}[k+(k-1)n]-\sum_{k=h}^{g-1}\left[N-k-(k-1)n\right]\right)^{2}\right.\\
\left.+\left(-\sum_{k=1}^{m}\sum_{i=1}^{n}(n-i+1)+\sum_{k=m+1}^{g-1}\sum_{i=1}^{n}(n-i+1)\right)^{2}\right\}\label{gz2poly-6}
\end{multline}
for $1\le h\le g$.
\begin{multline}
\left\langle {r_{Gh_{j}}^{2}}^{\hspace{-3mm}+}\right\rangle=\frac{l^{2}}{N^{2}}\left\{\left(\sum_{k=1}^{h-1}[k+(k-1)n]-\sum_{k=h}^{g-1}\left[N-k-(k-1)n\right]\right)^{2}\right.\\
\left.+\left(\sum_{i=1}^{j}(N-n+i-1)-\sum_{k=1}^{m}\sum_{i=1}^{n}(n-i+1)+\sum_{k=m+1}^{g-1}\sum_{i=1}^{n}(n-i+1)+\sum_{i=1}^{j}(n-i+1)\right)^{2}\right\}\label{gz2poly-7}
\end{multline}
for $2\le h\le m+1$.
\begin{multline}
\left\langle {r_{Gh_{j}}^{2}}^{\hspace{-3mm}-}\right\rangle=\frac{l^{2}}{N^{2}}\left\{\left(\sum_{k=1}^{h-1}[k+(k-1)n]-\sum_{k=h}^{g-1}\left[N-k-(k-1)n\right]\right)^{2}\right.\\
\left.+\left(-\sum_{i=1}^{j}(N-n+i-1)-\sum_{k=1}^{m}\sum_{i=1}^{n}(n-i+1)+\sum_{k=m+1}^{g-1}\sum_{i=1}^{n}(n-i+1)-\sum_{i=1}^{j}(n-i+1)\right)^{2}\right\}\label{gz2poly-8}
\end{multline}
for $m+2\le h\le g$.
\end{description}
By definition, the mean square of the radius of gyration can be calculated by the equation:
\begin{equation}
\left\langle s_{g}^{2}\right\rangle_{z=2}=\frac{1}{N}\left\{\sum_{h=1}^{g}\left\langle r_{Gh}^{2}\right\rangle+\sum_{h=2}^{m+1}\sum_{j=1}^{n}\left\langle {r_{Gh_{j}}^{2}}^{\hspace{-3mm}+}\right\rangle+\sum_{h=m+2}^{g}\sum_{j=1}^{n}\left\langle {r_{Gh_{j}}^{2}}^{\hspace{-3mm}-}\right\rangle\right\} \label{gz2poly-9}
\end{equation}
Let $r=\frac{m}{g-1}$ be the ratio of the side chains extending to the $Y(+)$ direction to the total number of the side chains. Since we are interested in the comb polymer having $\nu_{0}=1/4$, we put $n=g$ in Eqs. (\ref{gz2poly-6})-(\ref{gz2poly-9}). Eq. (\ref{gz2poly-9}) then gives the generalized expression of the fully expanded configuration:
\begin{equation}
\left\langle s_{g}^{2}\right\rangle_{z=2}=\frac{1}{6}\frac{(g^{2}-1)\left(g^{2}+3-6r+6g^{2}r+6r^{2}-6g^{2}r^{2}\right)}{g^{2}}\,l^{2} \label{gz2poly-10}
\end{equation}
where $0\le r=m/(g-1)\le 1$ as defined above. For a sufficiently large $g$, Eq. (\ref{gz2poly-10}) reduces to
\begin{equation}
\left\langle s_{g}^{2}\right\rangle_{z=2}=\frac{1}{6}\left(1+6r-6r^{2}\right)g^{2}l^{2}\label{gz2poly-11}
\end{equation}
Thus, $\left\langle s_{g}^{2}\right\rangle_{z=2}\propto g^{2}$. Since $N=g^{2}$ (Table \ref{MeanRGFreelyJointed}), this gives $\nu=1/2$. Within the generalized configuration illustrated in Fig. \ref{2D-z2}, there is no other exponent than $\nu=1/2$; only the pre-exponential factor, $A'=\frac{1}{6}(1+6r-6r^{2})$, changes: the factor, $A'$, changes from the minimum value, $1/6$ (at $r=0$ and $1$), to the maximum, $5/12$ (at $r=0.5$), as is seen in Fig. \ref{FullyExpanded-B}.

\begin{shaded}
\vspace*{-5mm}
\subsubsection*{Mathematical Check}
Consider the generalized expanded comb polymer with $g=3$ and $r=1/2$ on the square lattice. The monomers on this polymer can be allocated to the coordinates:
$$(-1,0), (0,0), (1,0), (0,1), (0,2), (0,3), (1,-1), (1,-2), (1,-3)$$
Following the elementary geometry, we have the center of masses, $\vec{r}_{OG}=\left(\frac{1}{3}, 0\right)$, from which we quickly find $\left\langle s_{g}^{2}\right\rangle_{z=2}=\frac{32}{9}\,l^{2}$, in exact agreement with the prediction of Eq. (\ref{gz2poly-10}).
\end{shaded}

\vspace*{5mm}
\section{Configuration of the Comb Polymer in the Three-dimensional Space}\label{CCP3D}
\begin{figure}[h]
\begin{center}
\includegraphics[width=13cm]{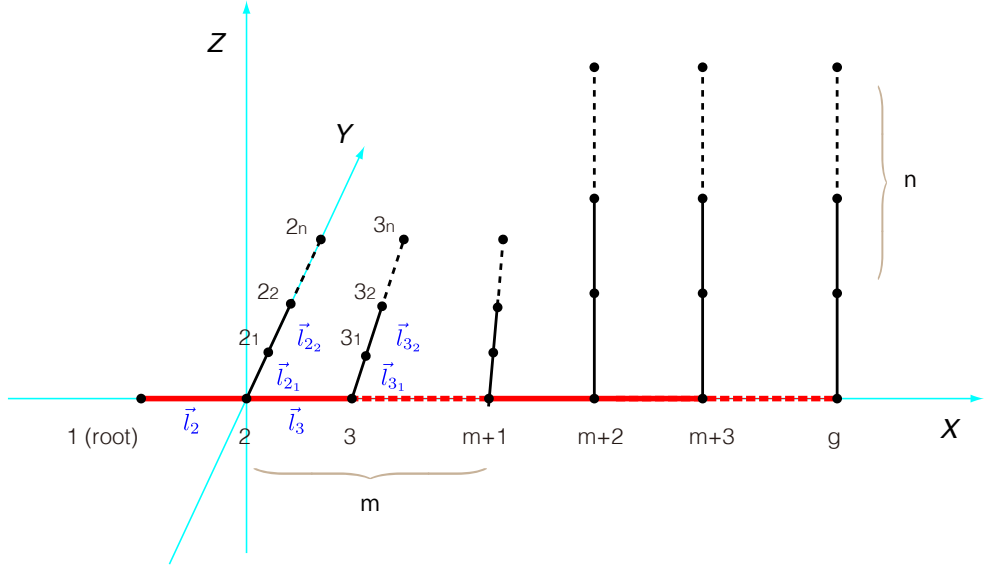} 
\vspace*{-7mm}
\caption{The comb polymer having the stretched backbone and the stretched side chains on the simple cubic lattice, with $m$ side chains extending on the $XY$ plane and $g-1-m$ on the $ZX$ plane.}\label{3D-z2}
\end{center}
\vspace*{-10mm}
\end{figure}
Let us consider another fully expanded configuration of the comb polymer in which the backbone and the side chains are in the stretched state, but $m$ of the $g-1$ side chains extend on the $XY$ plane and the remaining $g-1-m$ on the $ZX$ plane. In this configuration, the original vectorial expressions for the regular comb polymer\cite{Kazumi2} applies as it is. Only to divide the side chains into two groups is required: one is from $h=2$ to $m+1$, and the other is from $h=m+2$ to $g$. The resultant expressions for Eq. (\ref{gz2poly-2}) are
\begin{description}
\item[]
\begin{multline}
\vec{r}_{Gh}=\frac{1}{N}\left\{\sum_{k=1}^{h-1}[N-(N-k-(k-1)n)]\,\vec{l}_{(k+1)}-\sum_{k=h}^{g-1}\left[N-k-(k-1)n\right]\,\vec{l}_{(k+1)}\right.\\
\left.-\left(\sum_{k=1}^{m}\sum_{i=1}^{n}(n-i+1)\,\vec{l}_{(k+1)_{i}}\right)-\left(\sum_{k=m+1}^{g-1}\sum_{i=1}^{n}(n-i+1)\,\vec{l}_{(k+1)_{i}}\right)\right\}\label{gz2poly-12}
\end{multline}
for $1\le h\le g$.

\begin{multline}
\hspace{-0.5cm}\vec{r}_{Gh_{j}}=\frac{1}{N}\left\{\sum_{k=1}^{h-1}[N-(N-k-(k-1)n)]\,\vec{l}_{(k+1)}-\sum_{k=h}^{g-1}\left[N-k-(k-1)n\right]\,\vec{l}_{(k+1)}\right.\\
+\left(\sum_{i=1}^{j}[N-(n-i+1)]\,\vec{l}_{h_{i}}-\sum_{k=1}^{m}\sum_{i=1}^{n}(n-i+1)\,\vec{l}_{(k+1)_{i}}+\sum_{i=1}^{j}(n-i+1)\,\vec{l}_{h_{i}}\right)\\
\left.-\sum_{k=m+1}^{g-1}\sum_{i=1}^{n}(n-i+1)\,\vec{l}_{(k+1)_{i}}\right\}\label{gz2poly-13}
\end{multline}
for $2\le h\le m+1$.
\begin{multline}
\hspace{-0.5cm}\vec{r}_{Gh_{j}}=\frac{1}{N}\left\{\sum_{k=1}^{h-1}[N-(N-k-(k-1)n)]\,\vec{l}_{(k+1)}-\sum_{k=h}^{g-1}\left[N-k-(k-1)n\right]\,\vec{l}_{(k+1)}\right.\\
-\sum_{k=1}^{m}\sum_{i=1}^{n}(n-i+1)\,\vec{l}_{(k+1)_{i}}\\
\left.+\left(\sum_{i=1}^{j}[N-(n-i+1)]\,\vec{l}_{h_{i}}-\sum_{k=m+1}^{g-1}\sum_{i=1}^{n}(n-i+1)\,\vec{l}_{(k+1)_{i}}+\sum_{i=1}^{j}(n-i+1)\,\vec{l}_{h_{i}}\right)\right\}\label{gz2poly-14}
\end{multline}
for $m+2\le h\le g$.
\end{description}

\begin{figure}[h]
\vspace*{0mm}
\begin{center}
\includegraphics[width=9cm]{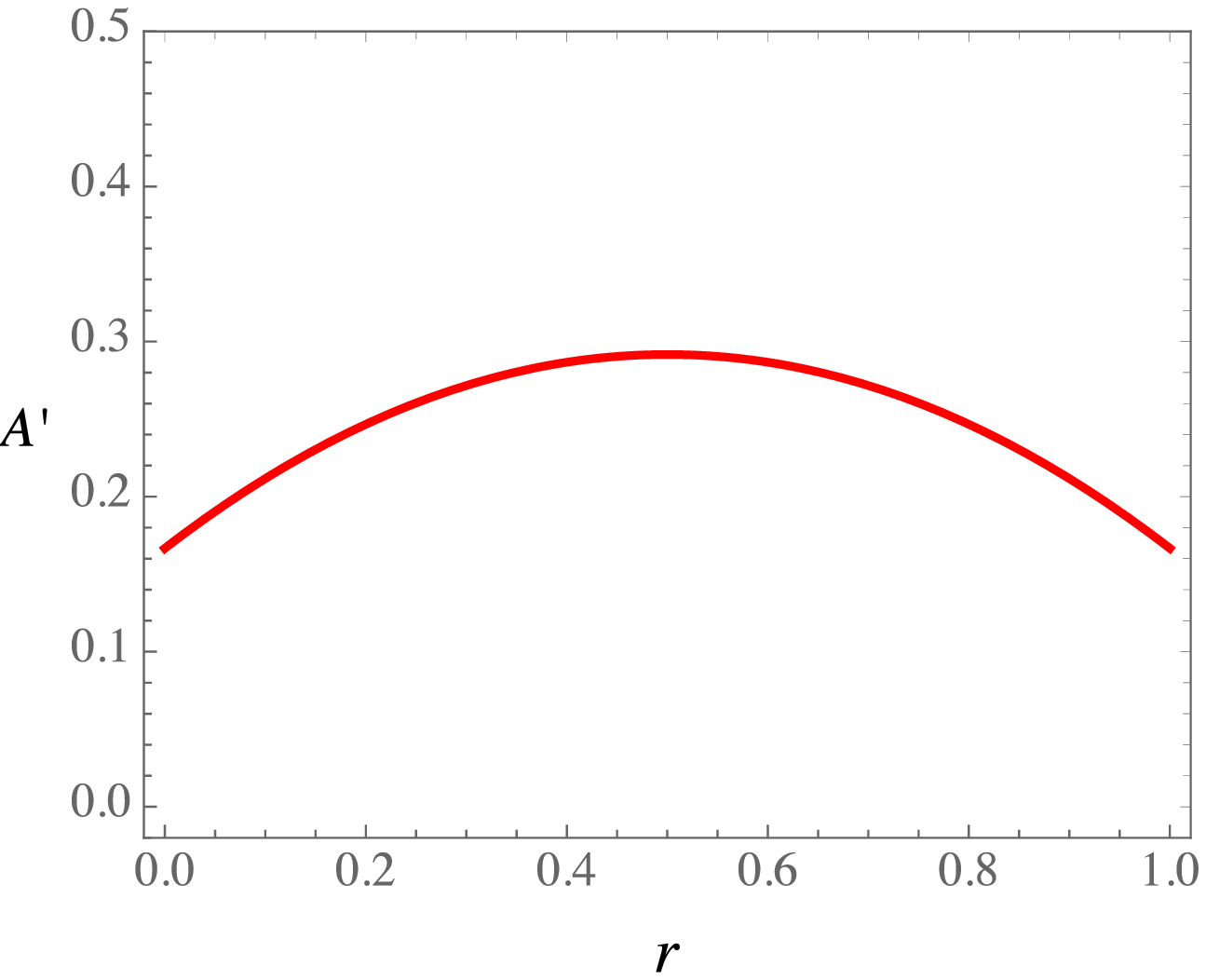} 
\caption{The prefactor, $A'=\frac{1}{6}\left(1+3r-3r^{2}\right)'$, as against the ratio, $r=\frac{m}{g-1}$, for the comb polymer having the stretched backbone and the stretched side chains on the simple cubic lattice, with $m$ side chains extending to $Y(+)$ and $g-1-m$ to $Z(+)$ (see Fig. \ref{3D-z2}).}\label{FullyExpanded-C}
\end{center}
 \vspace*{-3mm}
\end{figure}

The scalar products of bond vectors between the backbone and the side chains, and the corresponding product between the two side-chains groups, should vanish. Assuming the equal length, $|\vec{l}|=l$, for all bond vectors, the mean squares of the end-to-end distance can readily be calculated to yield
\vspace*{-1mm}
\begin{description}
\item[]
\begin{multline}
\left\langle r_{Gh}^{2}\right\rangle=\frac{l^{2}}{N^{2}}\left\{\left(\sum_{k=1}^{h-1}[k+(k-1)n]-\sum_{k=h}^{g-1}\left[N-k-(k-1)n\right]\right)^{2}\right.\\
\left.+\left(\sum_{k=1}^{m}\sum_{i=1}^{n}(n-i+1)\right)^{2}+\left(\sum_{k=m+1}^{g-1}\sum_{i=1}^{n}(n-i+1)\right)^{2}\right\}\label{gz2poly-15}
\end{multline}
for $1\le h\le g$.
\begin{multline}
\left\langle r_{Gh_{j}}^{2}\right\rangle=\frac{l^{2}}{N^{2}}\left\{\left(\sum_{k=1}^{h-1}[k+(k-1)n]-\sum_{k=h}^{g-1}\left[N-k-(k-1)n\right]\right)^{2}\right.\\
\left.+\left(\sum_{i=1}^{j}(N-n+i-1)-\sum_{k=1}^{m}\sum_{i=1}^{n}(n-i+1)+\sum_{i=1}^{j}(n-i+1)\right)^{2}+\left(\sum_{k=m+1}^{g-1}\sum_{i=1}^{n}(n-i+1)\right)^{2}\right\}\label{gz2poly-16}
\end{multline}
for $2\le h\le m+1$.
\begin{multline}
\left\langle r_{Gh_{j}}^{2}\right\rangle=\frac{l^{2}}{N^{2}}\left\{\left(\sum_{k=1}^{h-1}[k+(k-1)n]-\sum_{k=h}^{g-1}\left[N-k-(k-1)n\right]\right)^{2}\right.\\
\left.+\left(\sum_{k=1}^{m}\sum_{i=1}^{n}(n-i+1)\right)^{2}+\left(\sum_{i=1}^{j}(N-n+i-1)-\sum_{k=m+1}^{g-1}\sum_{i=1}^{n}(n-i+1)+\sum_{i=1}^{j}(n-i+1)\right)^{2}\right\}\label{gz2poly-17}
\end{multline}
for $m+2\le h\le g$,
\end{description}
Using the above results, the mean square of the radius of gyration can readily be calculated. Making the variable transformation, $n\rightarrow g$, we find
\begin{align}
\left\langle s_{g}^{2}\right\rangle_{z=2}&=\frac{1}{N}\left\{\sum_{h=1}^{g}\left\langle r_{Gh}^{2}\right\rangle+\sum_{h=2}^{m+1}\sum_{j=1}^{n}\left\langle r_{Gh_{j}}^{2}\right\rangle+\sum_{h=m+2}^{g}\sum_{j=1}^{n}\left\langle r_{Gh_{j}}^{2}\right\rangle\right\}\notag\\[2mm]
&=\frac{1}{6}\frac{(g^{2}-1)\left(g^{2}+3-3r+3g^{2}r+3r^{2}-3g^{2}r^{2}\right)}{g^{2}}\,l^{2} \hspace{5mm}(n=g)\label{gz2poly-18}
\end{align}
where $0\le r=m/(g-1)\le 1$. For a sufficiently large $g$, Eq. (\ref{gz2poly-18}) reduces to
\begin{equation}
\left\langle s_{g}^{2}\right\rangle_{z=2}=\frac{1}{6}\left(1+3r-3r^{2}\right)g^{2}l^{2}\label{gz2poly-19}
\end{equation}
Since $N=g^{2}$ by Table \ref{MeanRGFreelyJointed}, this again gives $\nu=1/2$, the same value as observed for the extended comb polymer. By simply changing the configuration of the side chains, one can not change the exponent, $\nu$; only the prefactor, $A'=\frac{1}{6}\left(1+3r-3r^{2}\right)$, changes from the minimum, $1/6$, to the maximum, $7/24$, (see Fig. \ref{FullyExpanded-C}).

The present result gives a proof that the stretched comb polymer in the three-dimensional space has the exponent, $\nu=1/2$, in agreement with the Issacson-Lubensky prediction, $\nu=\frac{2(1+\nu_{0})}{d+2}$, in $d=3$. The result reconfirms the sound basis of the thermodynamic theory of the excluded volume effects\cite{Flory, Issacson, Kazumi2}.

\begin{shaded}
\vspace*{-5mm}
\subsubsection*{Mathematical Check}
Let us consider the polymer with $g=3$ and $r=1/2$ in the three-dimensional space. The monomers on this polymer can be assigned to the following coordinates on the simple cubic lattice:
$$(-1,0,0), (0,0,0), (1,0,0), (0,1,0), (0,2,0), (0,3,0), (1,0,1), (1,0,2), (1,0,3)$$
According to the elementary geometry, we have the center of masses, $\vec{r}_{OG}=\left(\frac{1}{3}, \frac{2}{3}, \frac{2}{3}\right)$; from this we find $\left\langle s_{g}^{2}\right\rangle_{z=2}=\frac{8}{3}\,l^{2}$, in agreement with the prediction of Eq. (\ref{gz2poly-18}).
\end{shaded}

\section{Concluding Remarks}

For the fully expanded comb polymers, one can not change the exponent, $\nu$, by simply changing the configurations of the side chains, for instance, to the opposite directions (Fig. \ref{2D-z2}) or to the perpendicular directions (Fig. \ref{3D-z2}). By such configurational alteration, only the prefactor, $A'$, changes. As a result, the present results support the validity of the empirical equation:
\begin{equation}
\left\langle s_{g}^{2}\right\rangle=A'g^{2}l^{2} \hspace{5mm} (g\rightarrow\infty)\label{gz2poly-20}
\end{equation}
as a potential general rule for the fully stretched architectures. According to the argument in the preceding paper\cite{Kazumi2}, the relation (\ref{gz2poly-20}) is realizable for the isolated polymers that satisfy $\nu_{0}\le\frac{1}{d+1}$ in good solvents. These have the configurations that fulfill $\lambda=1$ and $\nu=2\nu_{0}$ as a result of the scaling relation, $\nu=2\lambda\nu_{0}$, where $\lambda$ is the exponent defined by $\left\langle s_{g}^{2}\right\rangle\propto g^{2\lambda}$. In Table \ref{MeanRGFullyExpanded}, the mean squares of the radii of gyration for various polymers with fully expanded configurations are shown. For all the cases, Eq. (\ref{gz2poly-20}) holds.

 \begin{table}[H]
 \vspace*{3mm}
 \centering
  \begin{threeparttable}
    \caption{The mean squares of the radii of gyration for architectures with fully expanded configurations $\left(r=\frac{m}{g-1}\right)$.}\label{MeanRGFullyExpanded}
\vspace*{-2mm}
  \begin{tabular}{l c c l}
\hline\\[-2mm]
\hspace{5mm} polymer  \hspace{0mm} &\hspace{-7mm} formulae for $\left\langle s_{g}^{2}\right\rangle$ &\hspace{-7mm} relationship between $N$ and $g$ &\hspace{3mm}  $\nu$\,\,\,\,\\[2mm]
\hline\\[-1.5mm]
\hspace{5mm} linear  & $\big(\frac{1}{12}\big)\,(g^{2}-1)\, l^{2}$\hspace{7 mm} &\hspace{-6mm} $N=g$ &\hspace{3mm}  $1$\\[1.5mm]
\hspace{5mm} $z$=$2$\tnote{\,a}  &$\big(\frac{1}{6}\big)\frac{(g^{2}-1)(g^{2}+3)}{g^{2}}\, l^{2}$\hspace{7 mm} &\hspace{-5mm}  $N=g^{2}$ &\hspace{3mm}  $\frac{1}{2}$\\[1.5mm]
\hspace{5mm} extended comb I\tnote{\,b}  &$\big(\frac{1}{6}\big)\frac{(g^{2}-1)\left(g^{2}+3-6r+6g^{2}r+6r^{2}-6g^{2}r^{2}\right)}{g^{2}}\,l^{2}$\hspace{7 mm} &\hspace{-5mm}  $N=g^{2}$ &\hspace{3mm}  $\frac{1}{2}$\\[1.5mm]
\hspace{5mm} extended comb \II\tnote{\,c}  &$\big(\frac{1}{6}\big)\frac{(g^{2}-1)\left(g^{2}+3-3r+3g^{2}r+3r^{2}-3g^{2}r^{2}\right)}{g^{2}}\,l^{2}$\hspace{7 mm} &\hspace{-5mm}  $N=g^{2}$ &\hspace{3mm}  $\frac{1}{2}$\\[1.5mm]
\hspace{5mm} randomly branched  &\rule[2pt]{10mm}{0.5pt}\hspace{10 mm} &\hspace{-5mm} \hspace{0mm}\rule[2pt]{1cm}{0.5pt} &\hspace{3mm}  \hspace{0mm}\rule[2pt]{0.3cm}{0.5pt}\\[1.5mm]
\hspace{5mm} $z$=3\tnote{\,d}  &$\big(\frac{1}{12}\big)\frac{(g-1)(3g^{5}-5g^{4}+16g^{3}+6g^{2}+g-3)}{(g^{2}-g+1)^{2}}\, l^{2}$\hspace{7 mm} &\hspace{-5mm}  $N=g(g^{2}-g+1)$ &\hspace{3mm}  $\frac{1}{3}$\\[1.5mm]
\hline\\[-6mm]
   \end{tabular}
   \vspace*{1mm}
   \begin{tablenotes}
     \item a. the nested architecture with the depth $z$=2 (equivalent to the extended comb polymer with $n=g$)\cite{Kazumi2}; b. see Fig. \ref{2D-z2}; c. see Fig. \ref{3D-z2}; d. the nested architecture with the depth $z$=3\cite{Kazumi2}.
   \end{tablenotes}
  \end{threeparttable}
  \vspace*{3mm}
\end{table}

The configurational isomers investigated in the present paper, by no means, cover all possible stretched configurations. Notwithstanding, we can state conclusively that the extended comb polymer I with $\nu_{0}=1/4$ takes the stretched configuration of $\lambda=1$ in the two-dimensional space ($d=2$), so that $\nu=1/2$. This is because of the reason that the fully stretched configuration is the maximum state of the expansion of the polymer, and the calculation in Section \ref{CCP2D} has yielded $\nu=\frac{1}{2}$, so that we must have $\nu\le\frac{1}{2}$, whereas the critical packing density criterion requires $\nu\ge 1/d_{c}=\frac{1}{2}$; hence $\nu=\frac{1}{2}$ and $\lambda=1$ are confirmed.

\vspace*{8mm}

\end{document}